\documentclass[5p]{elsarticle}

\pdfoutput=1

\usepackage{graphicx}

\title{Adaptive mutation of biochemical reaction constants:\\ Fisher's geometrical model without pleiotropy}

\author[rng]{Ryan N. Gutenkunst\corref{cor1}} \ead{ryan@gutenkunst.org}
\author[jps]{James P. Sethna} \ead{sethna@lassp.cornell.edu}
\cortext[cor1]{Corresponding author}
\address[rng]{Theoretical Biology and Biophysics \& Center for Nonlinear Studies, Los Alamos National Laboratory, Los Alamos, NM 87545}
\address[jps]{Laboratory of Atomic and Solid State Physics, Cornell University, Ithaca, NY 14853}

\usepackage{xspace}
\usepackage{amsmath}
\usepackage{amssymb}
\usepackage[usenames,dvipsnames]{color}
\newcommand{\transpose}[1]{\ensuremath{{#1}^\mathsf{T}}}
\newcommand{\mat}[1]{\ensuremath{\mathbf{#1}}}
\newcommand{\rhati}{\ensuremath{\hat{r}_i}\xspace}
\newcommand{\erf}{\operatorname{erf}}
\newcommand{\smalldot}{\ensuremath{\!\cdot\!}}

\begin{document}

\begin{abstract}
The distribution of fitness effects of adaptive mutations remains poorly understood, both empirically and theoretically.
We study this distribution using a version of Fisher's geometrical model without pleiotropy, such that each mutation affects only a single trait.
We are motivated by the notion of an organism's chemotype, the set of biochemical reaction constants that govern its molecular constituents.
From physical considerations, we expect the chemotype to be of high dimension and to exhibit very little pleiotropy.
Our model generically predicts striking cusps in the distribution of the fitness effects of arising and fixed mutations.
It further predicts that a single element of the chemotype should comprise all mutations at the high-fitness ends of these distributions.
Using extreme value theory, we show that the two cusps with the highest fitnesses are typically well-separated, even when the chemotype possesses thousands of elements; this suggests a means to observe these cusps experimentally.
More broadly, our work demonstrates that new insights into evolution can arise from the chemotype perspective, a perspective between the genotype and the phenotype.
\end{abstract}

\begin{keyword}
fitness landscape; pleiotropy; fitness effects distribution; extreme value theory
\end{keyword}

\maketitle

\section{Introduction}

Adaptive mutation is fundamental to the evolutionary process, and it is medically important to the emergence of drug resistance in microbes~\cite{bib:Woodford2007} and tumors~\cite{bib:Merlo2006}.
Given the selective advantage of a mutation, the probability that it fixes in a population (i.e., rises to frequency 1) and the mean time to do so are well-known~\cite{bib:Ewens2004}.
Comparatively little is known, however, about the distribution of selective advantages among new mutations.
This distribution can be experimentally measured by confronting genetically identical populations with a novel environment such a new food source and measuring the fitness of newly arising mutations~\cite{bib:Elena2003}.
Such measurements are difficult, because adaptive mutations are rare; thus theoretical analysis can offer important insights~\cite{bib:Orr2005}.


A popular predictive framework for studying adaptive evolution is R.\ A.\ Fisher's geometrical model, which considers adaptation in phenotypic ``trait'' space~\cite{bib:Fisher1930}.
Mutations are characterized by the phenotypic changes they induce, which correspond to moves in trait space.
Fisher used this model to argue that evolution is primarily driven by the accumulation of many mutations that each have only a small effect~\cite{bib:Fisher1930}.
This argument was influential until Motoo Kimura pointed out that mutations with larger effects are more likely to fix, so most adaptive mutations that fix have intermediate effect~\cite{bib:Kimura1983}.

Recent studies have applied Fisher's model to a gamut of questions in evolutionary biology and population genetics; these include the distribution of mutation fitness effects near an optimum~\cite{bib:Martin2008}, sequential adaptation~\cite{bib:Orr1998,bib:Orr1999}, and the load of deleterious mutations carried by finite populations~\cite{bib:Hartl1998,bib:Poon2000}.
Of particular note, predictions from the model regarding epistasis compare favorably with data~\cite{bib:Martin2007}.
The model predicts a roughly exponential distribution of fitness effects for new mutations~\cite{bib:Orr2006}, similar to mutational landscape models of adaptive evolution~\cite{bib:Orr2002}.
This prediction is consistent with experiments in viruses~\cite{bib:Rokyta2005} and bacteria~\cite{bib:Kassen2006}, although more recent experiments by Rokyta et al.\ point toward a truncated distribution~\cite{bib:Rokyta2008}.
Here we consider a geometrical model without pleiotropy, a model in which each mutation affects only a single trait.
We are motivated by considering the phenotype at a finer scale than is typical.


One can view the information specifying an organism through a variety of scales~\cite{bib:Daniels2008}.
On the largest scale, the phenotype of the entire organism, a single mutation often affects multiple traits, implying substantial pleiotropy.
On the finest scale, the genotype, a single mutation often affects only one amino acid codon or one regulatory binding site, implying no pleiotropy.
Systems biology is often modeled at the intermediate scale of biochemical reaction constants; multiple codons combine to determine a single biochemical reaction constant and multiple constants combine to determine a single phenotypic trait.
Motivated by this useful intermediate level of description, we introduced the word ``chemotype''~\cite{bib:Daniels2008} to refer to the set of biochemical reaction constants determining the rates of molecular reactions in an organism.
Other authors have considered specific biochemical reaction constants to be aspects of the phenotype, for example Hartl, Dykhuizen and Dean~\cite{bib:Hartl1985}.
We find it useful to distinguish the chemotype, because it differs in important ways from the large-scale phenotype typically considered.


The chemotype differs from the large-scale phenotype in both dimensionality and pleiotropy.
The number of independent high-level phenotypic traits for even a complex organism may be modest~\cite{bib:Martin2006,bib:Tenaillon2007}.
The number of independent elements of a chemotype, however, is comparable to the number of an organism's genes.
Each gene codes for a protein or RNA with its own biochemical reaction constants, so each gene contributes at least one element to the chemotype.
The chemotype is additionally distinguished by very low pleiotropy, the degree to which single mutations affect multiple traits.
Recent experiments on mouse skeletal traits have demonstrated that this system possess a moderate degree of pleiotropy; a given mutation typically affects around five traits~\cite{bib:Wagner2008}.
By contrast, a single mutation is expected to affect only one or a few elements of the chemotype.
This is because single-nucleotide mutations are dominant in short-term and laboratory evolution~\cite{bib:Gresham2006,bib:Lynch2008}, and they typically change only a single protein residue or a single DNA binding site.
Such a change will in turn impact only one or a few biochemical reaction constants, implying very low pleiotropy in chemotype space.


Other authors have considered zero pleiotropy geometric models in the study of drift load~\cite{bib:Peck1997,bib:Poon2000}.
We focus here on the distributions of fitness effects of adaptive mutations that arise and that subsequently fix in a population.
A general argument shows that these distributions possess sharp cusps, one for each element of the chemotype.
Given the high dimensionality of chemotype space, however, it is not obvious whether these cusps are observable.
To address this question, we study a more specific model, in which the fitness landscape is Gaussian.
For this model, we show using extreme value theory that the two cusps with the highest fitnesses are well-separated, even in a space with thousands of dimensions.
This suggests that the cusps, and thus the nature of evolution in chemotype space, can be studied experimentally.

\section{Model}

\begin{figure}
\centering
\includegraphics{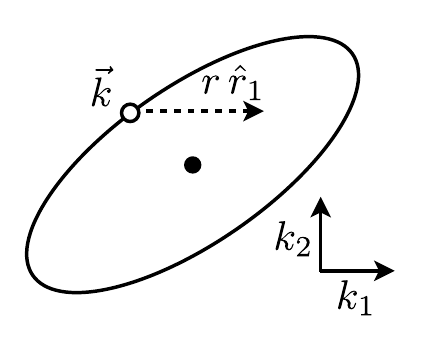}
\caption{
Illustration of the model.
Motived by the evolution of biochemical reaction constants, we consider evolution in a high-dimensional space with no pleiotropy (chemotype space).
The current uniform chemotype $\vec{k}$ of a population is a point in this space.
The optimal chemotype is the origin of our coordinate system and lies at the center of the fitness contours.
In the absence of pleiotropy, mutations change one element of $\vec{k}$ at a time, so moves are made along the coordinate axes.
The dashed arrow indicates an adaptive mutation of magnitude $r$ in element $k_1$.\label{fig:evo:model}}
\end{figure}

As illustrated in Fig.~\ref{fig:evo:model}, the state of an organism with $N$ chemotype elements can be represented as a point in $N$-dimensional space: $\vec{k} = \left<k_1, k_2, \dots k_N\right>$.
The change in state caused by a mutation can be described by an $N$-dimensional vector $\vec{r}$; the mutant has chemotype $\vec{k} + \vec{r}$.
Because mutations will typically change only one or a few elements of the chemotype, most pairs of mutations are \emph{orthogonal} in this space.
We thus restrict our attention to mutations with zero pleiotropy, which change only a single chemotype element at a time.
Thus $\vec{r} = r\,\rhati$, where $r$ the size of the mutation (which may be negative) and $\rhati$ is a unit vector along the $i$th coordinate axis. 
We define each $\rhati$ so that mutations which increase fitness have positive values of $r$.
We work in the limit of strong selection and weak mutation, so that the population is genetically homogenous aside from rare mutants that arise one at a time and either fix or are lost before the next mutation arises.
In this limit, the state of the entire population corresponds to a single point $\vec{k}$ in chemotype space, and fixation of the mutation $\vec{r}$ moves the entire population to $\vec{k} + \vec{r}$.

\subsection{Gaussian landscape}

For analysis, we specialize to a Gaussian fitness landscape in which the fitness $W(\vec{k})$ of a population with chemotype $\vec{k}$ is
\begin{align}\label{eqn:gauss}
W(\vec{k}) = \exp\left(-\frac{1}{2} \, \vec{k} \smalldot \mat{S} \smalldot \vec{k}\right),
\end{align}
where $\mat{S}$ is a symmetric positive definite matrix and $\mat{S} \smalldot \vec{k}$ denotes the dot product of the matrix $\mat{S}$ and the vector $\vec{k}$.
Without loss of generality, the optimum fitness is set to one.

It is convenient to work with the logarithmic fitness change $Q$, introduced by Waxman and Welch~\cite{bib:Waxman2005} and defined as
\begin{align}
Q &\equiv \log\left[W(\vec{k} + \vec{r})\, \big/ \, W(\vec{k})\right].
\end{align}
$Q$ is related to the selection coefficient $s$ by $s = e^Q - 1$, and for mutations with small selective advantage $Q \approx s$. 
Adaptive mutations are those with $Q > 0$.
For the Gaussian landscape, the log-fitness change caused by a mutation of size $r$ in chemotype element $i$ is
\begin{align}
Q_i(r) = -\vec{k} \smalldot \mat{S} \smalldot \rhati\, r - \frac{1}{2}\, \rhati \smalldot \mat{S} \smalldot \rhati\, r^2.\label{eqn:evo:Qexpanded}
\end{align}
The largest possible gain in log-fitness achievable by mutating chemotype element $i$ is denoted $\theta_i$ and obtained by maximizing $Q_i(r)$ with respect to $r$:
\begin{align}
\theta_i &= \frac{\left|\vec{k}\smalldot\mat{S}\smalldot\rhati\right|^2}{2 \,\rhati\smalldot\mat{S}\smalldot\rhati}\label{eqn:evo:zeta}.
\end{align}
The magnitude of the largest possible mutation in chemotype element $i$ that can be made without decreasing fitness is $\rho_i$:
\begin{align}
\rho_i = 2 \frac{\left|\vec{k}\smalldot\mat{S}\smalldot\rhati\right|} {\rhati\smalldot\mat{S}\smalldot\rhati}.
\end{align}
These quantities are illustrated in Fig.~\ref{fig:defs}A.

Many of our results are derived for spherically symmetric fitness landscapes, for which $\mat{S} = \lambda \mat{I}$, where $\mat{I}$ is the identity matrix.
For such a landscape,
\begin{align}
\theta_i = \frac{\lambda | \vec{k} \smalldot \rhati | ^2}{2} \equiv Q_{tot} |\hat{k} \smalldot \rhati|^2 \label{eqn:evo:zetasphere},
\end{align}
and
\begin{align}
\rho_i &= 2 |\vec{k} \smalldot \rhati|.\label{eqn:evo:rhosym}
\end{align}
Here $Q_{tot} \equiv \lambda |\vec{k}|/2 = -\log W(\vec{k})$ is the difference in log-fitness between the optimum (which has a fitness of one) and the current chemotype.



\section{Results}

We first show generally that adaptation on a smooth fitness landscape without pleiotropy leads to cusps in the distribution of fitness effects of newly arising and fixed mutations, one cusp for each element of the chemotype (i.e., each dimension of the landscape).
To assess whether these cusps will be observable, we specialize to Gaussian landscapes for which we can perform detailed calculations.
We show that the cusps will be difficult to directly observe in a histogram of experimental fitness measurements.
Nevertheless, we show that the cusps do have observable consequences, because the two cusps with the highest fitnesses are typically well-separated, even if the landscape has thousands of dimensions and deviates strongly from a sphere.

\subsection{Cusps}

\begin{figure*}
\centering
\includegraphics{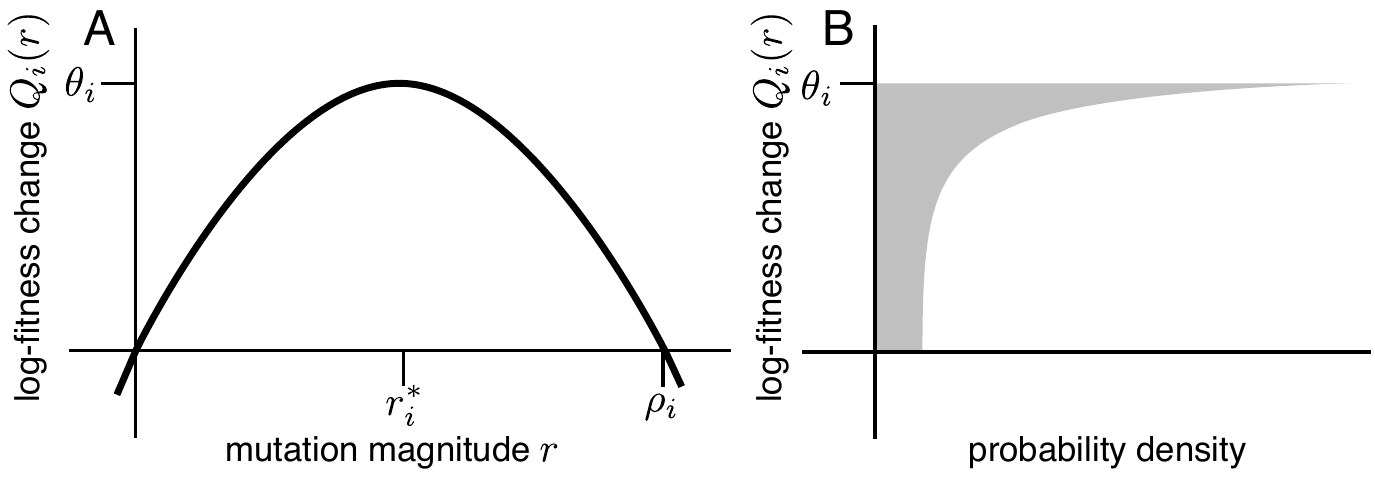}
\caption{
Origin of cusps. A: The log-fitness $Q_i(r)$ as a function of the mutation size $r$ in chemotype element $i$. $Q_i(r)$ has a maximum of $\theta_i$ at $r = r_i^*$, and it returns to zero at $r = \rho_i$.
B: The probability density of mutation fitness effects. It generically has a cusp at $Q_i(r) = \theta_i$, corresponding to the point at which $Q_i(r)$ has zero slope. The density is plotted sideways to emphasize the connection between the cusp and the maximum of $Q_i(r)$.
\label{fig:defs}}
\end{figure*}

Fig.~\ref{fig:defs}A illustrates the slice $Q_i$ of the fitness landscape accessible by mutations to a particular chemotype element $i$.
For a mutation of size $r$, the range of mutations $\Delta r$ about $r$ that produce fitness changes in a given range $\Delta Q$ is inversely proportional to the slope $\mathrm{d}Q_i/\mathrm{d}r$.
By definition, $Q_i$ is at a maximum for the fittest mutation $r_i^*$ and thus has zero slope there.
This yields an infinite inverse and thus a cusp in the distribution of fitness effects of adaptive mutations to chemotype element $i$, as illustrated in Fig.~\ref{fig:defs}B.

The above argument relies only on a lack of pleiotropy, a smooth fitness landscape, and a distribution of fitness affects which is non-singular and broad enough to access the optimal mutation of a given chemotype element.
A natural question is whether these cusps will be observable in experiments.
To address this, we analyze a more specific landscape for which we can perform concrete calculations.

\subsection{Gaussian landscape}

Comparisons between empirical mutation effect distributions in different environments for several organisms support a Gaussian form for the fitness landscape close to the optimum chemotype~\cite{bib:Martin2006a}.
For the remainder of this paper, we assume a Gaussian landscape as in Eq.~\ref{eqn:gauss}.

We must also specify the distribution of mutational effects $f(r)$ on the chemotype.
In ~\ref{sec:mutation_scale}, we use the fact that most mutations are deleterious~\cite{bib:Eyre-Walker2007,bib:Perfeito2007} to show that this distribution must typically span the full range of adaptive mutations for all elements of the chemotype.
In other words, mutations must often `hop over' the ridge of increased fitness.
For computational ease, we take $f(r)$ to be uniform over the range of adaptive mutations; our qualitative results are robust to this choice.

\begin{figure}
\centering
\includegraphics{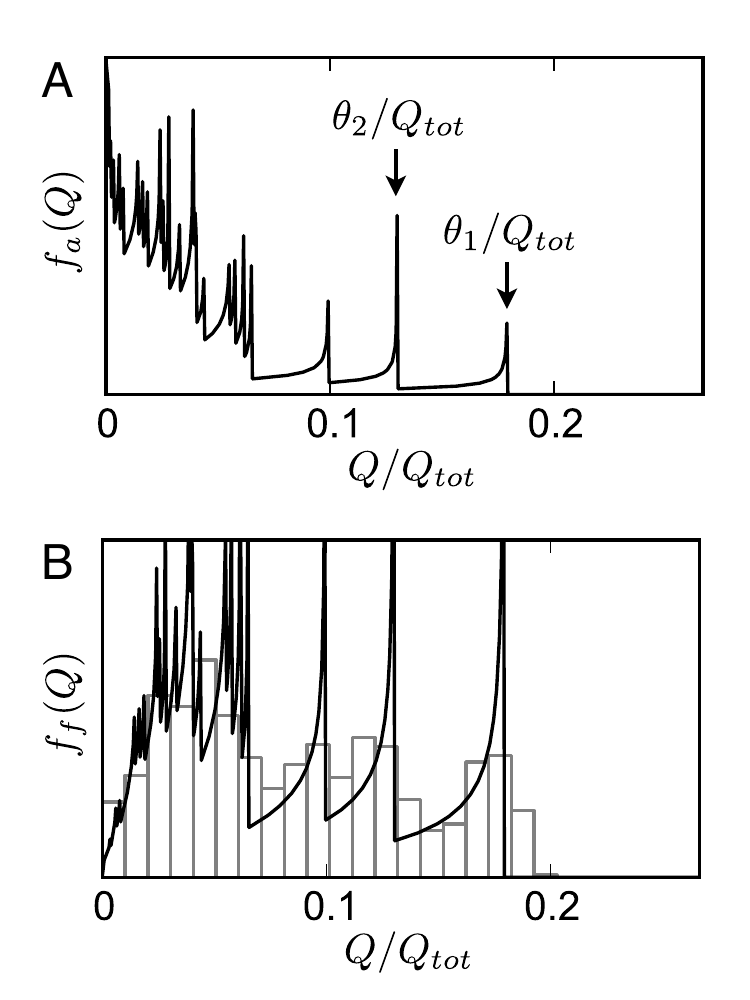}
\caption{
Example of cusps. A: The probability distribution $f_a(Q)$ of the fitness effects of adaptive mutations for a 30-dimensional spherical fitness landscape with a random initial chemotype $\vec{k}$. 
B:~The probability distribution $f_f(Q)$ of fitness effects of fixed mutations, for the same $\vec{k}$ as in A. The cusps at large values of the fitness $Q$ are much more prominent. The grey histogram shows 1000 samples from the distribution, each including a 1\% error in the measurement of $Q/Q_{tot}$.
\label{fig:evo:cusps}}
\end{figure}

Given these two assumptions, we can calculate the distribution $f_a(Q)$ of fitness effects for adaptive mutations, as detailed in~\ref{sec:distribution}:
\begin{align}
f_{a}(Q) 
 & \propto \sum_i \frac{1}{\sqrt{\rhati\smalldot\mat{S}\smalldot\rhati} \sqrt{\theta_i - Q}} 
 \label{eqn:evo:faq}.
\end{align}
Here the sum is over elements of the chemotype, and each element contributes its own cusp.
Fig.~\ref{fig:evo:cusps}A plots $f_a(Q)$ for a population at a particular random chemotype $\vec{k}$ in a 30-dimensional spherical fitness landscape (in which $\mat{S}$ is proportional to the identity matrix).
Note that the distribution is bound by a roughly exponential envelope.
As detailed in~\ref{sec:average}, if we average our distribution over initial states $\vec{k}$ with a given fitness, we obtain Waxman and Welch's previous result~\cite{bib:Waxman2005} for the spherical model with maximum pleiotropy.

For a large population, the probability that an adaptive mutation fixes is proportional to its fitness effect $Q$~\cite{bib:Kimura1983,bib:Sella2005}; thus the density of fitness effects of fixed mutations $f_f(Q)$ is
\begin{equation}
f_f(Q) \propto Q f_a(Q).\label{eqn:evo:ff}
\end{equation}
This density of fixed mutations is shown in Fig.~\ref{fig:evo:cusps}B for the same spherical fitness landscape and initial chemotype $\vec{k}$ as in Fig.~\ref{fig:evo:cusps}A.
The cusps at large $Q$ are much more prominent in the distribution of fixed mutation fitness effects.

Distributions of fitness effects can be measured experimentally by introducing identical populations to identical novel environments and tracking mutations that sweep through them.
The histogram in Fig.~\ref{fig:evo:cusps}B simulates such an experiment, representing 1000 samples from $f_f(Q)$, each polluted by Gaussian measurement noise in $Q/Q_{tot}$ with a standard deviation 0.01.
Given the rarity of adaptive mutations, such an experiment would be difficult, but even such a difficult experiment cannot directly resolve the cusps.
We now show, however, that the spacing between the fittest and second-fittest cusps is typically substantial.
This implies that the upper end of the fitness distribution should be dominated by mutations to a single element of the chemotype, a prediction that can be tested experimentally.


\subsubsection{Cusp spacings}
Each cusp in Fig.~\ref{fig:evo:cusps} corresponds to mutations affecting a different chemotype element $k_i$.
Thus our model not only predicts cusps, but also predicts that the most adaptive mutations will all affect the \emph{same} element of the chemotype.
To experimentally test this prediction, it suffices to measure relative fitness differences of order $\Delta$, where
\begin{equation}
\Delta \equiv (\theta_1 - \theta_2)/\theta_1
\end{equation}
is the normalized separation between the two cusps with the highest fitnesses.
In ~\ref{sec:evo:appevt} we derive the distribution of $\Delta$ predicted by our model for a spherical landscape, using methods of extreme value theory~\cite{bib:Gumbel1958}.

\begin{figure}
\centering
\includegraphics{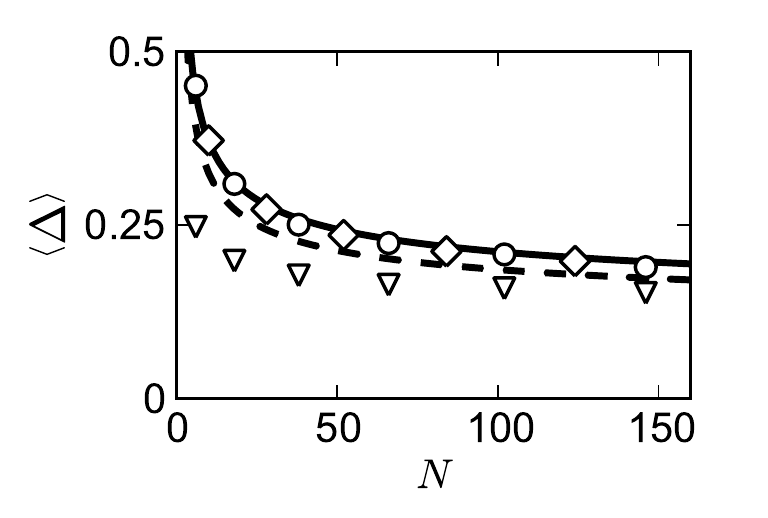}
\caption{
Mean relative spacing $\Delta$ between the fittest and second-fittest cusps as a function of landscape dimension $N$.
The solid line is the asymptotically exact result for the spherical fitness landscape, while the dashed line is the approximation of Eq.~\ref{eqn:evo:approx}.
The circles are numerical simulations for the spherical landscape, while the diamonds and triangles are simulation results for mildly and severely non-spherical landscapes, respectively.
The mean value of $\Delta$ declines very slowly with $N$ (especially for large $N$), suggesting that the cusps will typically be well-separated even for chemotype spaces of high dimension. \label{fig:evo:cusplocs}}
\end{figure}

The solid line in Fig.~\ref{fig:evo:cusplocs} is the exact asymptotic result (using Eq.~\ref{eqn:evo:u} and~\ref{eqn:evo:alpha}) for the mean of $\Delta$, given a spherical fitness landscape.
The dashed line is the approximation
\begin{equation}
\left<\Delta\right> \approx \frac{1}{1 + \log N + \log \sqrt{2/\pi}},\label{eqn:evo:approx}
\end{equation}
which is valid for large landscape dimension $N$.
The circles in Fig.~\ref{fig:evo:cusplocs} are the results from numerical simulations in the spherical landscape at each $N$. 
The agreement between the exact asymptotic result and the numerical simulations is excellent, and the approximate result captures the trend well.
Note that $\left<\Delta\right>$ declines very slowly as a function of $N$; for a chemotype with $N = 10,000$ elements the mean $\Delta$ is approximately 0.11, a relative fitness difference that is straightforward to measure experimentally.
For comparison, Fig.~\ref{fig:evo:cusps} has $\Delta \approx 0.27$, which is approximately the predicted $\left<\Delta\right>$ when the chemotype has $N = 30$ elements.
Thus our model predicts that, even for a high-dimensional chemotype, a substantial range $\Delta$ of the most adaptive mutations will affect the same element of the chemotype.

\subsubsection{Non-spherical landscapes}\label{sec:evo:nonspherical}
To assess the robustness of our results regarding cusp spacing, we numerically test them in non-spherical landscapes. 
Studying non-spherical fitness landscapes also implicitly considers different mutation scales amongst chemotype elements, because any differences in the typical size of chemotype mutation effects on different elements (anisotropic $f(r)$) can be eliminated by rescaling the chemotype elements $k_i$.
A convenient way to characterize a landscape is by the eigenvalues of $\mat{S}$.
Spherical landscapes have all eigenvalues equal, while for non-spherical landscapes the width of the fitness contour along any given eigenvector of $\mat{S}$ is proportional to the square root of the corresponding eigenvalue, so landscapes with a larger range of eigenvalues are more non-spherical.

For a given landscape $\mat{S}$ and initial chemotype $\vec{k}$, $\Delta$  can be calculated numerically from the definition of $\theta_i$. 
In the tests described below, the mean of $\Delta$ is calculated from $10^4$ simulations, each instance involving an independent landscape $\mat{S}$ and initial chemotype $\vec{k}$.
The eigenvectors of $\mat{S}$ were random orthogonal vectors, and the initial chemotypes were chosen at random among those with a fixed fitness $Q_{tot}$, as described in \ref{sec:numerical_sampling}.

The diamonds in Fig.~\ref{fig:evo:cusplocs} result from mildly non-spherical fitness landscapes corresponding to eigenvalues of $\mat{S}$ drawn uniformly from the range $0.4<\lambda<3.6$, following Waxman~\cite{bib:Waxman2007}.
The deviations of $\left<\Delta\right>$ from the spherical case are small.

The triangles in Fig.~\ref{fig:evo:cusplocs} arise from ``sloppy'' fitness landscapes~\cite{bib:Brown2003a,bib:Gutenkunst2007} with the $N$  eigenvalues evenly spaced in the logarithm from $10^6$ to $10^{-6}$.
This corresponds to the narrowest axis of the fitness contours being one-millionth the width of the longest axis.
Even for these very non-spherical fitness landscapes, the average spacing $\left<\Delta\right>$ remains substantial and comparable to the average in the spherical case.

\section{Discussion}
We analyzed adaptive mutation in a version of Fisher's geometric model in which mutations are restricted to acting in only one dimension at a time.
This condition of zero pleiotropy is appropriate when the population is described in terms of its chemotype, the biochemical reaction constants of the molecules that comprise the organism, only one or a few of which will be altered by any given point mutation.
We showed that the probability density of fitness effects of adaptive mutations will generically exhibit cusps, each associated with mutations of a particular chemotype element.
These cusps are particularly prominent in the density of fitness effects of fixed mutations.
Simulations suggest that directly resolving these cusps experimentally will be difficult.
However, each cusp corresponds to a different element of the chemotype, and we showed that the relative spacing between the two cusps with the highest fitness remains substantial even in very non-spherical landscapes of high dimension.
This suggests a testable prediction that is robust to details of the model: the fittest mutations should all affect the same element of the chemotype.

It may be surprising that even very non-spherical fitness landscapes (range in eigenvalues of $10^{12}$) yield a qualitatively similar cusp distribution to the spherical landscape.
Our simulations assume that the eigenvectors, and thus the correlations between chemotype elements and fitness, are random.
In this case, each chemotype element contributes about equally to each eigenvector, so the fitness function is similar when projected along each chemotype direction, yielding a narrow distribution of $\theta_i$ which is similar to the spherical case.
The assumption of random correlation structure is motivated by empirical study of the sensitivity of biochemical networks to reaction constant variation~\cite{bib:Gutenkunst2007} and theoretical study of sloppy systems in general~\cite{bib:Waterfall2006}.
In both cases, random eigenvectors are a reasonable approximation to the complicated correlations found.

A key assumption of our model is that each chemotype element is continuously adjustable throughout the range of possible adaptive mutations.
Because the genetic code is discrete, this cannot be strictly true.
The distribution of effects of random mutations on chemotype elements is not well-known, in part because most biochemical studies focus on mutations of large effect.
However, studies have shown that random mutations can introduce small but non-zero changes to the enzymatic activity of proteins~\cite{bib:Bloom2007} and the expression driven by promoter sites~\cite{bib:Landry2007}, suggesting that our assumption of continuous chemotype variation is reasonable.



The substantial average separation $\left<\Delta\right>$ we predict between the fittest and second-fittest cusp implies that the mutations conveying the largest fitness benefits will all involve a single chemotype element.
A similar result holds for the mutational landscape model~\cite{bib:Orr2002,bib:Orr2003}, in which the largest fitness spacings are between the fittest sequences.
The spacing distributions in the mutational landscape model, however, depend on the correlation assumed between the effects of different mutations.
A recent analysis of such correlations considers mutations within ``blocks'' of sequence that contribute independently and identically to fitness~\cite{bib:Orr2006a}.
Each block may be roughly interpreted as a different chemotype element in our model, but in our model the relative contributions to fitness differ between blocks and naturally arise from the structure of the landscape.
Nevertheless, the fact that both models predict the upper end of the fitness distribution to be dominated by few mutations, or in our case mutations in a few chemotype elements, may help explain the large amount of parallel evolution that can be observed in separate populations exposed to similar environments~\cite{bib:Wood2005,bib:Pelosi2006}.

The distribution of fitness effects of adaptive mutations has been studied in bacteria and viruses~\cite{bib:Imhof2001,bib:Kassen2006}.
Typically the distribution is found to be consistent with a smooth exponential distribution.
Our theory predicts only gentle cusps in this distribution, but much more prominent cusps in the distribution of fitness of effects of fixed mutations.
This distribution has been studied experimentally in bacteria~\cite{bib:Rozen2002,bib:Barrett2006,bib:Perfeito2007}, and those results are also consistent with a smooth distribution.
We showed, however, that it would be difficult for such experiments to directly resolve the cusps.
Intriguingly, a recent study of virus adaptation by Rokyta et al.\ points toward a fitness effects distribution with a truncated right-hand tail~\cite{bib:Rokyta2008}, consistent with our model.


Directly resolving the cusps is challenging; it will be easier to test the prediction that the fittest mutations will all affect a single element of the chemotype.
Given that the average fittest cusp separation $\left<\Delta\right>$ is roughly 0.1 even for very large $N$, resolving this effect requires a relative precision in fitness of a few percent, which is achievable by averaging repeated assays.
Recent developments in microarray-based genotyping~\cite{bib:Gresham2006} allow the sites of mutations to be cost-effectively identified.
Mutations that reside in, for example, the same region of a protein likely affect the same element of the chemotype.
Correlating fitness measurements of mutations with identification of which chemotype element they affect will allow direct testing of our model predictions.

Motivated by the adaptation of the chemotype, the set of biochemical rate constants comprising an organism, we have studied a version of Fisher's geometrical model without pleiotropy.
The model  predicts cusps in the distribution of fitness effects of fixed mutations and that the fittest mutations all involve a single element of the chemotype.
Analysis suggests that the second prediction is experimentally accessible.
More broadly, our work suggests that viewing evolution in terms of the chemotype may offer new insights beyond those found at the genotype or phenotype level.

\section*{Acknowledgments}
We thank Carl Franck, whose exam question prompted this investigation. We also thank Jason Mezey and Ben Logsdon for helpful discussions relating to population genetics and evolution and Chris Myers, Josh Waterfall, and Fergal Casey for discussions of the model itself.
This work was supported by US National Science Foundation grant DMR-070167.

\bibliography{database}

\begin{thebibliography}{46}
\expandafter\ifx\csname natexlab\endcsname\relax\def\natexlab#1{#1}\fi
\providecommand{\bibinfo}[2]{#2}
\ifx\xfnm\relax \def\xfnm[#1]{\unskip,\space#1}\fi
\bibitem[{Woodford and Ellington(2007)}]{bib:Woodford2007}
\bibinfo{author}{N.~Woodford}, \bibinfo{author}{M.~J. Ellington},
\newblock \bibinfo{title}{The emergence of antibiotic resistance by mutation},
\newblock \bibinfo{journal}{Clin. Microbiol. Infect.} \bibinfo{volume}{13}
  (\bibinfo{year}{2007}) \bibinfo{pages}{5--18}.
\bibitem[{Merlo et~al.(2006)Merlo, Pepper, Reid, and Maley}]{bib:Merlo2006}
\bibinfo{author}{L.~M.~F. Merlo}, \bibinfo{author}{J.~W. Pepper},
  \bibinfo{author}{B.~J. Reid}, \bibinfo{author}{C.~C. Maley},
\newblock \bibinfo{title}{Cancer as an evolutionary and ecological process},
\newblock \bibinfo{journal}{Nat. Rev. Cancer} \bibinfo{volume}{6}
  (\bibinfo{year}{2006}) \bibinfo{pages}{924--935}.
\bibitem[{Ewens(2004)}]{bib:Ewens2004}
\bibinfo{author}{W.~J. Ewens}, \bibinfo{title}{Mathematical Population
  Genetics}, volume~\bibinfo{volume}{1}, \bibinfo{publisher}{Springer-Verlag},
  \bibinfo{edition}{2nd} edition, \bibinfo{year}{2004}.
\bibitem[{Elena and Lenski(2003)}]{bib:Elena2003}
\bibinfo{author}{S.~F. Elena}, \bibinfo{author}{R.~E. Lenski},
\newblock \bibinfo{title}{Evolution experiments with microorganisms: the
  dynamics and genetic bases of adaptation},
\newblock \bibinfo{journal}{Nat. Rev. Genet.} \bibinfo{volume}{4}
  (\bibinfo{year}{2003}) \bibinfo{pages}{457--469}.
\bibitem[{Orr(2005)}]{bib:Orr2005}
\bibinfo{author}{H.~A. Orr},
\newblock \bibinfo{title}{Theories of adaptation: what they do and don't say},
\newblock \bibinfo{journal}{Genetica} \bibinfo{volume}{123}
  (\bibinfo{year}{2005}) \bibinfo{pages}{3--13}.
\bibitem[{Fisher(1930)}]{bib:Fisher1930}
\bibinfo{author}{R.~A. Fisher}, \bibinfo{title}{The genetical theory of natural
  selection}, \bibinfo{publisher}{Oxford Univ Press}, \bibinfo{address}{Oxford,
  U.K.}, \bibinfo{year}{1930}.
\bibitem[{Kimura(1983)}]{bib:Kimura1983}
\bibinfo{author}{M.~Kimura}, \bibinfo{title}{The neutral theory of molecular
  evolution}, \bibinfo{publisher}{Cambridge Univ. Press},
  \bibinfo{address}{Cambridge, U.K.}, \bibinfo{year}{1983}.
\bibitem[{Martin and Lenormand(2008)}]{bib:Martin2008}
\bibinfo{author}{G.~Martin}, \bibinfo{author}{T.~Lenormand},
\newblock \bibinfo{title}{The distribution of beneficial and fixed mutation
  fitness effects close to an optimum},
\newblock \bibinfo{journal}{Genetics} \bibinfo{volume}{179}
  (\bibinfo{year}{2008}) \bibinfo{pages}{907--916}.
\bibitem[{Orr(1998)}]{bib:Orr1998}
\bibinfo{author}{H.~A. Orr},
\newblock \bibinfo{title}{The population genetics of adaptation: the
  distribution of factors fixed during adaptive evolution},
\newblock \bibinfo{journal}{Evolution} \bibinfo{volume}{52}
  (\bibinfo{year}{1998}) \bibinfo{pages}{935--949}.
\bibitem[{Orr(1999)}]{bib:Orr1999}
\bibinfo{author}{H.~A. Orr},
\newblock \bibinfo{title}{The evolutionary genetics of adaptation: a simulation
  study},
\newblock \bibinfo{journal}{Genet. Res.} \bibinfo{volume}{74}
  (\bibinfo{year}{1999}) \bibinfo{pages}{207--214}.
\bibitem[{Hartl and Taubes(1998)}]{bib:Hartl1998}
\bibinfo{author}{D.~L. Hartl}, \bibinfo{author}{C.~H. Taubes},
\newblock \bibinfo{title}{Towards a theory of evolutionary adaptation},
\newblock \bibinfo{journal}{Genetica} \bibinfo{volume}{102-103}
  (\bibinfo{year}{1998}) \bibinfo{pages}{525--533}.
\bibitem[{Poon and Otto(2000)}]{bib:Poon2000}
\bibinfo{author}{A.~Poon}, \bibinfo{author}{S.~P. Otto},
\newblock \bibinfo{title}{Compensating for our load of mutations: freezing the
  meltdown of small populations},
\newblock \bibinfo{journal}{Evolution} \bibinfo{volume}{54}
  (\bibinfo{year}{2000}) \bibinfo{pages}{1467--1479}.
\bibitem[{Martin et~al.(2007)Martin, Elena, and Lenormand}]{bib:Martin2007}
\bibinfo{author}{G.~Martin}, \bibinfo{author}{S.~F. Elena},
  \bibinfo{author}{T.~Lenormand},
\newblock \bibinfo{title}{Distributions of epistasis in microbes fit
  predictions from a fitness landscape model},
\newblock \bibinfo{journal}{Nat. Genet.} \bibinfo{volume}{39}
  (\bibinfo{year}{2007}) \bibinfo{pages}{555--560}.
\bibitem[{Orr(2006)}]{bib:Orr2006}
\bibinfo{author}{H.~A. Orr},
\newblock \bibinfo{title}{The distribution of fitness effects among beneficial
  mutations in {F}isher's geometric model of adaptation},
\newblock \bibinfo{journal}{J. Theor. Biol.} \bibinfo{volume}{238}
  (\bibinfo{year}{2006}) \bibinfo{pages}{279--285}.
\bibitem[{Orr(2002)}]{bib:Orr2002}
\bibinfo{author}{H.~A. Orr},
\newblock \bibinfo{title}{The population genetics of adaptation: the adaptation
  of {DNA} sequences},
\newblock \bibinfo{journal}{Evolution} \bibinfo{volume}{56}
  (\bibinfo{year}{2002}) \bibinfo{pages}{1317--1330}.
\bibitem[{Rokyta et~al.(2005)Rokyta, Joyce, Caudle, and
  Wichman}]{bib:Rokyta2005}
\bibinfo{author}{D.~R. Rokyta}, \bibinfo{author}{P.~Joyce},
  \bibinfo{author}{S.~B. Caudle}, \bibinfo{author}{H.~A. Wichman},
\newblock \bibinfo{title}{An empirical test of the mutational landscape model
  of adaptation using a single-stranded {DNA} virus},
\newblock \bibinfo{journal}{Nat. Genet.} \bibinfo{volume}{37}
  (\bibinfo{year}{2005}) \bibinfo{pages}{441--444}.
\bibitem[{Kassen and Bataillon(2006)}]{bib:Kassen2006}
\bibinfo{author}{R.~Kassen}, \bibinfo{author}{T.~Bataillon},
\newblock \bibinfo{title}{Distribution of fitness effects among beneficial
  mutations before selection in experimental populations of bacteria},
\newblock \bibinfo{journal}{Nat. Genet.} \bibinfo{volume}{38}
  (\bibinfo{year}{2006}) \bibinfo{pages}{484--488}.
\bibitem[{Rokyta et~al.(2008)Rokyta, Beisel, Joyce, Ferris, Burch, and
  Wichman}]{bib:Rokyta2008}
\bibinfo{author}{D.~R. Rokyta}, \bibinfo{author}{C.~J. Beisel},
  \bibinfo{author}{P.~Joyce}, \bibinfo{author}{M.~T. Ferris},
  \bibinfo{author}{C.~L. Burch}, \bibinfo{author}{H.~A. Wichman},
\newblock \bibinfo{title}{Beneficial fitness effects are not exponential for
  two viruses},
\newblock \bibinfo{journal}{J. Molec. Evol.} \bibinfo{volume}{67}
  (\bibinfo{year}{2008}) \bibinfo{pages}{368--376}.
\bibitem[{Daniels et~al.(2008)Daniels, Chen, Sethna, Gutenkunst, and
  Myers}]{bib:Daniels2008}
\bibinfo{author}{B.~C. Daniels}, \bibinfo{author}{Y.-J. Chen},
  \bibinfo{author}{J.~P. Sethna}, \bibinfo{author}{R.~N. Gutenkunst},
  \bibinfo{author}{C.~R. Myers},
\newblock \bibinfo{title}{Sloppiness, robustness, and evolvability in systems
  biology},
\newblock \bibinfo{journal}{Curr. Opin. Biotech.} \bibinfo{volume}{19}
  (\bibinfo{year}{2008}) \bibinfo{pages}{389 -- 395}.
\bibitem[{Hartl et~al.(1985)Hartl, Dykhuizen, and Dean}]{bib:Hartl1985}
\bibinfo{author}{D.~L. Hartl}, \bibinfo{author}{D.~E. Dykhuizen},
  \bibinfo{author}{A.~M. Dean},
\newblock \bibinfo{title}{Limits of adaptation: the evolution of selective
  neutrality},
\newblock \bibinfo{journal}{Genetics} \bibinfo{volume}{111}
  (\bibinfo{year}{1985}) \bibinfo{pages}{655--674}.
\bibitem[{Martin and Lenormand(2006)}]{bib:Martin2006}
\bibinfo{author}{G.~Martin}, \bibinfo{author}{T.~Lenormand},
\newblock \bibinfo{title}{A general multivariate extension of {F}isher's
  geometrical model and the distribution of mutation fitness effects across
  species},
\newblock \bibinfo{journal}{Evolution} \bibinfo{volume}{60}
  (\bibinfo{year}{2006}) \bibinfo{pages}{893--907}.
\bibitem[{Tenaillon et~al.(2007)Tenaillon, Silander, Uzan, and
  Chao}]{bib:Tenaillon2007}
\bibinfo{author}{O.~Tenaillon}, \bibinfo{author}{O.~K. Silander},
  \bibinfo{author}{J.-P. Uzan}, \bibinfo{author}{L.~Chao},
\newblock \bibinfo{title}{Quantifying organismal complexity using a population
  genetic approach},
\newblock \bibinfo{journal}{PLoS ONE} \bibinfo{volume}{2}
  (\bibinfo{year}{2007}) \bibinfo{pages}{e217}.
\bibitem[{Wagner et~al.(2008)Wagner, Kenney-Hunt, Pavlicev, Peck, Waxman, and
  Cheverud}]{bib:Wagner2008}
\bibinfo{author}{G.~P. Wagner}, \bibinfo{author}{J.~P. Kenney-Hunt},
  \bibinfo{author}{M.~Pavlicev}, \bibinfo{author}{J.~R. Peck},
  \bibinfo{author}{D.~Waxman}, \bibinfo{author}{J.~M. Cheverud},
\newblock \bibinfo{title}{Pleiotropic scaling of gene effects and the `cost of
  complexity'},
\newblock \bibinfo{journal}{Nature} \bibinfo{volume}{452}
  (\bibinfo{year}{2008}) \bibinfo{pages}{470--472}.
\bibitem[{Gresham et~al.(2006)Gresham, Ruderfer, Pratt, Schacherer, Dunham,
  Botstein, and Kruglyak}]{bib:Gresham2006}
\bibinfo{author}{D.~Gresham}, \bibinfo{author}{D.~M. Ruderfer},
  \bibinfo{author}{S.~C. Pratt}, \bibinfo{author}{J.~Schacherer},
  \bibinfo{author}{M.~J. Dunham}, \bibinfo{author}{D.~Botstein},
  \bibinfo{author}{L.~Kruglyak},
\newblock \bibinfo{title}{Genome-wide detection of polymorphisms at nucleotide
  resolution with a single {DNA} microarray},
\newblock \bibinfo{journal}{Science} \bibinfo{volume}{311}
  (\bibinfo{year}{2006}) \bibinfo{pages}{1932--1936}.
\bibitem[{Lynch et~al.(2008)Lynch, Sung, Morris, Coffey, Landry, Dopman,
  Dickinson, Okamoto, Kulkarni, Hartl, and Thomas}]{bib:Lynch2008}
\bibinfo{author}{M.~Lynch}, \bibinfo{author}{W.~Sung},
  \bibinfo{author}{K.~Morris}, \bibinfo{author}{N.~Coffey},
  \bibinfo{author}{C.~R. Landry}, \bibinfo{author}{E.~B. Dopman},
  \bibinfo{author}{W.~J. Dickinson}, \bibinfo{author}{K.~Okamoto},
  \bibinfo{author}{S.~Kulkarni}, \bibinfo{author}{D.~L. Hartl},
  \bibinfo{author}{W.~K. Thomas},
\newblock \bibinfo{title}{A genome-wide view of the spectrum of spontaneous
  mutations in yeast},
\newblock \bibinfo{journal}{Proc Natl Acad Sci U S A} \bibinfo{volume}{105}
  (\bibinfo{year}{2008}) \bibinfo{pages}{9272--9277}.
\bibitem[{Peck et~al.(1997)Peck, Barreau, and Heath}]{bib:Peck1997}
\bibinfo{author}{J.~R. Peck}, \bibinfo{author}{G.~Barreau},
  \bibinfo{author}{S.~C. Heath},
\newblock \bibinfo{title}{Imperfect genes, {F}isherian mutation and the
  evolution of sex},
\newblock \bibinfo{journal}{Genetics} \bibinfo{volume}{145}
  (\bibinfo{year}{1997}) \bibinfo{pages}{1171--1199}.
\bibitem[{Waxman and Welch(2005)}]{bib:Waxman2005}
\bibinfo{author}{D.~Waxman}, \bibinfo{author}{J.~J. Welch},
\newblock \bibinfo{title}{Fisher's microscope and {H}aldane's ellipse},
\newblock \bibinfo{journal}{Am. Nat.} \bibinfo{volume}{166}
  (\bibinfo{year}{2005}) \bibinfo{pages}{447--457}.
\bibitem[{Martin and Lenormand(2006)}]{bib:Martin2006a}
\bibinfo{author}{G.~Martin}, \bibinfo{author}{T.~Lenormand},
\newblock \bibinfo{title}{The fitness effect of mutations across environments:
  a survey in light of fitness landscape models},
\newblock \bibinfo{journal}{Evolution} \bibinfo{volume}{60}
  (\bibinfo{year}{2006}) \bibinfo{pages}{2413--2427}.
\bibitem[{Eyre-Walker and Keightley(2007)}]{bib:Eyre-Walker2007}
\bibinfo{author}{A.~Eyre-Walker}, \bibinfo{author}{P.~D. Keightley},
\newblock \bibinfo{title}{The distribution of fitness effects of new
  mutations},
\newblock \bibinfo{journal}{Nat. Rev. Genet.} \bibinfo{volume}{8}
  (\bibinfo{year}{2007}) \bibinfo{pages}{610--618}.
\bibitem[{Perfeito et~al.(2007)Perfeito, Fernandes, Mota, and
  Gordo}]{bib:Perfeito2007}
\bibinfo{author}{L.~Perfeito}, \bibinfo{author}{L.~Fernandes},
  \bibinfo{author}{C.~Mota}, \bibinfo{author}{I.~Gordo},
\newblock \bibinfo{title}{Adaptive mutations in bacteria: high rate and small
  effects},
\newblock \bibinfo{journal}{Science} \bibinfo{volume}{317}
  (\bibinfo{year}{2007}) \bibinfo{pages}{813--815}.
\bibitem[{Sella and Hirsh(2005)}]{bib:Sella2005}
\bibinfo{author}{G.~Sella}, \bibinfo{author}{A.~E. Hirsh},
\newblock \bibinfo{title}{The application of statistical physics to
  evolutionary biology},
\newblock \bibinfo{journal}{Proc. Natl. Acad. Sci. USA} \bibinfo{volume}{102}
  (\bibinfo{year}{2005}) \bibinfo{pages}{9541--9546}.
\bibitem[{Gumbel(1958)}]{bib:Gumbel1958}
\bibinfo{author}{E.~J. Gumbel}, \bibinfo{title}{Statistics of Extremes},
  \bibinfo{publisher}{Columbia University Press}, \bibinfo{address}{New York},
  \bibinfo{year}{1958}.
\bibitem[{Waxman(2007)}]{bib:Waxman2007}
\bibinfo{author}{D.~Waxman},
\newblock \bibinfo{title}{Mean curvature versus normality: a comparison of two
  approximations of {F}isher's geometrical model},
\newblock \bibinfo{journal}{Theor. Popul. Biol.} \bibinfo{volume}{71}
  (\bibinfo{year}{2007}) \bibinfo{pages}{30--36}.
\bibitem[{Brown and Sethna(2003)}]{bib:Brown2003a}
\bibinfo{author}{K.~S. Brown}, \bibinfo{author}{J.~P. Sethna},
\newblock \bibinfo{title}{Statistical mechanical approaches to models with many
  poorly known parameters},
\newblock \bibinfo{journal}{Phys. Rev. E} \bibinfo{volume}{68}
  (\bibinfo{year}{2003}) \bibinfo{pages}{021904}.
\bibitem[{Gutenkunst et~al.(2007)Gutenkunst, Waterfall, Casey, Brown, Myers,
  and Sethna}]{bib:Gutenkunst2007}
\bibinfo{author}{R.~N. Gutenkunst}, \bibinfo{author}{J.~J. Waterfall},
  \bibinfo{author}{F.~P. Casey}, \bibinfo{author}{K.~S. Brown},
  \bibinfo{author}{C.~R. Myers}, \bibinfo{author}{J.~P. Sethna},
\newblock \bibinfo{title}{Universally sloppy parameter sensitivities in systems
  biology},
\newblock \bibinfo{journal}{PLoS Comput. Biol.} \bibinfo{volume}{3}
  (\bibinfo{year}{2007}) \bibinfo{pages}{e189}.
\bibitem[{Waterfall et~al.(2006)Waterfall, Casey, Gutenkunst, Brown, Myers,
  Brouwer, Elser, and Sethna}]{bib:Waterfall2006}
\bibinfo{author}{J.~J. Waterfall}, \bibinfo{author}{F.~P. Casey},
  \bibinfo{author}{R.~N. Gutenkunst}, \bibinfo{author}{K.~S. Brown},
  \bibinfo{author}{C.~R. Myers}, \bibinfo{author}{P.~W. Brouwer},
  \bibinfo{author}{V.~Elser}, \bibinfo{author}{J.~P. Sethna},
\newblock \bibinfo{title}{Sloppy-model universality class and the {V}andermonde
  matrix},
\newblock \bibinfo{journal}{Phys. Rev. Lett.} \bibinfo{volume}{97}
  (\bibinfo{year}{2006}) \bibinfo{pages}{150601}.
\bibitem[{Bloom et~al.(2007)Bloom, Romero, Lu, and Arnold}]{bib:Bloom2007}
\bibinfo{author}{J.~D. Bloom}, \bibinfo{author}{P.~A. Romero},
  \bibinfo{author}{Z.~Lu}, \bibinfo{author}{F.~H. Arnold},
\newblock \bibinfo{title}{Neutral genetic drift can alter promiscuous protein
  functions, potentially aiding functional evolution},
\newblock \bibinfo{journal}{Biol. Direct} \bibinfo{volume}{2}
  (\bibinfo{year}{2007}) \bibinfo{pages}{17}.
\bibitem[{Landry et~al.(2007)Landry, Lemos, Rifkin, Dickinson, and
  Hartl}]{bib:Landry2007}
\bibinfo{author}{C.~R. Landry}, \bibinfo{author}{B.~Lemos},
  \bibinfo{author}{S.~A. Rifkin}, \bibinfo{author}{W.~J. Dickinson},
  \bibinfo{author}{D.~L. Hartl},
\newblock \bibinfo{title}{Genetic properties influencing the evolvability of
  gene expression},
\newblock \bibinfo{journal}{Science} \bibinfo{volume}{317}
  (\bibinfo{year}{2007}) \bibinfo{pages}{118--121}.
\bibitem[{Orr(2003)}]{bib:Orr2003}
\bibinfo{author}{H.~A. Orr},
\newblock \bibinfo{title}{A minimum on the mean number of steps taken in
  adaptive walks},
\newblock \bibinfo{journal}{J. Theor. Biol.} \bibinfo{volume}{220}
  (\bibinfo{year}{2003}) \bibinfo{pages}{241--247}.
\bibitem[{Orr(2006)}]{bib:Orr2006a}
\bibinfo{author}{H.~A. Orr},
\newblock \bibinfo{title}{The population genetics of adaptation on correlated
  fitness landscapes: the block model},
\newblock \bibinfo{journal}{Evolution} \bibinfo{volume}{60}
  (\bibinfo{year}{2006}) \bibinfo{pages}{1113--1124}.
\bibitem[{Wood et~al.(2005)Wood, Burke, and Rieseberg}]{bib:Wood2005}
\bibinfo{author}{T.~E. Wood}, \bibinfo{author}{J.~M. Burke},
  \bibinfo{author}{L.~H. Rieseberg},
\newblock \bibinfo{title}{Parallel genotypic adaptation: when evolution repeats
  itself},
\newblock \bibinfo{journal}{Genetica} \bibinfo{volume}{123}
  (\bibinfo{year}{2005}) \bibinfo{pages}{157--170}.
\bibitem[{Pelosi et~al.(2006)Pelosi, Kühn, Guetta, Garin, Geiselmann, Lenski,
  and Schneider}]{bib:Pelosi2006}
\bibinfo{author}{L.~Pelosi}, \bibinfo{author}{L.~Kühn},
  \bibinfo{author}{D.~Guetta}, \bibinfo{author}{J.~Garin},
  \bibinfo{author}{J.~Geiselmann}, \bibinfo{author}{R.~E. Lenski},
  \bibinfo{author}{D.~Schneider},
\newblock \bibinfo{title}{Parallel changes in global protein profiles during
  long-term experimental evolution in {E}scherichia coli},
\newblock \bibinfo{journal}{Genetics} \bibinfo{volume}{173}
  (\bibinfo{year}{2006}) \bibinfo{pages}{1851--1869}.
\bibitem[{Imhof and Schlotterer(2001)}]{bib:Imhof2001}
\bibinfo{author}{M.~Imhof}, \bibinfo{author}{C.~Schlotterer},
\newblock \bibinfo{title}{Fitness effects of advantageous mutations in evolving
  escherichia coli populations},
\newblock \bibinfo{journal}{Proc. Natl. Acad. Sci. USA} \bibinfo{volume}{98}
  (\bibinfo{year}{2001}) \bibinfo{pages}{1113--1117}.
\bibitem[{Rozen et~al.(2002)Rozen, de~Visser, and Gerrish}]{bib:Rozen2002}
\bibinfo{author}{D.~E. Rozen}, \bibinfo{author}{J.~A. G.~M. de~Visser},
  \bibinfo{author}{P.~J. Gerrish},
\newblock \bibinfo{title}{Fitness effects of fixed beneficial mutations in
  microbial populations},
\newblock \bibinfo{journal}{Curr. Biol.} \bibinfo{volume}{12}
  (\bibinfo{year}{2002}) \bibinfo{pages}{1040--1045}.
\bibitem[{Barrett et~al.(2006)Barrett, MacLean, and Bell}]{bib:Barrett2006}
\bibinfo{author}{R.~D.~H. Barrett}, \bibinfo{author}{R.~C. MacLean},
  \bibinfo{author}{G.~Bell},
\newblock \bibinfo{title}{Mutations of intermediate effect are responsible for
  adaptation in evolving {P}seudomonas fluorescens populations},
\newblock \bibinfo{journal}{Biol. Lett.} \bibinfo{volume}{2}
  (\bibinfo{year}{2006}) \bibinfo{pages}{236--238}.
\bibitem[{Abramowitz and Stegun(1964)}]{bib:Abramowitz1964}
\bibinfo{author}{M.~Abramowitz}, \bibinfo{author}{I.~A. Stegun},
  \bibinfo{title}{Handbook of Mathematical Functions with Formulas, Graphs, and
  Mathematical Tables}, \bibinfo{publisher}{Dover}, \bibinfo{address}{New
  York}, \bibinfo{edition}{10th} edition, \bibinfo{year}{1964}.

\end{thebibliography}

\appendix


\section{Scale of chemotype mutation}\label{sec:mutation_scale}


If the distribution $f(r)$ of mutation effects on the chemotype were small for $r$ greater than the typical $\rho_i$, the mutational distance over which adaptive mutations are possible, then the fraction $P_a$ of mutations that were adaptive would be approximately one-half.
The rarity of adaptive mutations thus suggests that $f(r)$ must be appreciable for $r$ greater than the typical $\rho_i$.
We now show that $f(r)$ must remain substantial even for $r$ greater than the \emph{largest} $\rho_i$.
To do so, we make the simplifying assumption that the distribution of mutational effects is identical for all chemotype elements.
We then consider the scenario in which this distribution barely covers the range of all possible adaptive mutations, extending only to the largest of the $\rho_i$, demoted  $\max_i \rho_i$.
For this scenario, we derive an analytic approximation to $P_a$ for spherical landscapes, and we calculate $P_a$ numerically for the non-spherical landscapes considered in Fig.~\ref{fig:evo:cusplocs}.
In both cases we find that this scenario leads to an unrealistically high probability of adaptive mutation, implying that the distribution of mutation chemotype effects must have a scale larger than that of the largest possible adaptive mutation.

If the probability density of mutation chemotype effects $f(r)$ were uniform over $(-\max_i \rho_i, +\max_i \rho_i)$, the probability of a random mutation being adaptive would be
\begin{align}
P_a &= \frac{\sum_i \rho_i}{2 N \max_i \rho_i}.
\end{align}
The numerator is the total length of intervals where mutations are adaptive, and the denominator is the total length of intervals over which mutations are distributed.

Specializing to the spherical case and plugging in for $\rho_i$, we have
\begin{align}
P_a &= \frac{\left<|\hat{k} \smalldot \rhati|\right>}{2 \max_i | \hat{k} \smalldot \rhati|}\label{eqn:adapt_sphere}.
\end{align}
Asymptotically for large $N$, $\hat{k} \smalldot \rhati$ has a Gaussian probability density with variance $1/N$.
Averaging yields
\begin{equation}
\left<|\hat{k} \smalldot \rhati |\right>~=~\sqrt{2}/\sqrt{\pi N}.
\end{equation}
The largest absolute value of $N$ samples drawn from a Gaussian density with variance $1/N$ is asymptotically $\sqrt{\left.{2 \log \left(N/\sqrt{2\pi}\right)}\right/{N}}$~\cite[Eq.~4.2.3(11)]{bib:Gumbel1958}.
Plugging these into Eq.~\ref{eqn:adapt_sphere} yields
\begin{equation}
P_a\left(N\right) \sim \frac{1}{2 \sqrt{\pi \log(N/\sqrt{2 \pi})}}.\label{eqn:evo:Pben}
\end{equation}
This probability remains substantial even for very large $N$. For example, $P_a\left(10,000\right)$ is roughly 0.1, an unrealistically large value.

Numerical tests with both the mildly and wildly non-spherical landscapes considered in Fig.~\ref{fig:evo:cusplocs} yield a $P_a$ of at least 0.13, consistent with the spherical result of Eq.~\ref{eqn:evo:Pben}.
This suggests that, for a realistic fraction of mutations to be deleterious, the typical scale of chemotype effects for mutations must be larger than $\max_i \rho_i$, even for very non-spherical landscapes.

\section{Fitness effects distribution}\label{sec:distribution}
Given the probability density of chemotype mutation effects $f(r)$, the distribution $f_a(Q)$ of fitness effects for adaptive mutations is
\begin{equation}
f_{a}(Q) \propto \sum_i \int dr f(r) \,\delta(Q - Q_i(r)),
\end{equation}
where the sum is over all chemotype elements $i$, and $Q_i(r)$ is given by Eq.~\ref{eqn:evo:Qexpanded}.
Making the variable substitution $u = Q_i(r)$ yields
\begin{align}
f_{a}(Q) & \propto \sum_i \int du \frac{f(Q_i^{-1}(u)) \,\delta(Q - u)}{\sqrt{\left|\vec{k}\smalldot\mat{S}\smalldot\rhati\right|^2- 2 \, \rhati\smalldot\mat{S}\smalldot\rhati \, u}}. 
\end{align}
Further making the approximation that $f(r)$ is uniform over the range of adaptive mutations, and substituting Eq.~\ref{eqn:evo:zeta} yields Eq.~\ref{eqn:evo:faq}.

\section{Average distributions}\label{sec:average}

The distribution of adaptive fitness effects averaged over initial chemotypes with a given fitness can be calculated by averaging $f_{a}(Q)$ over the probability density for $\theta_i$.
For a spherical fitness landscape, the $\theta_i$ are proportional to the squared magnitudes of the components of the unit vector $\hat{k}$.
Asymptotically as the number of dimensions $N \rightarrow \infty$, these are squares of Gaussian variables and have probability density
\begin{equation}
f_\theta(\theta_i) \propto \exp{\left[-\theta_i N/(2 Q_{tot})\right]}/\sqrt{\theta_i},\label{eqn:evo:chi2}
\end{equation}
which is a $\chi^2$ density with one degree of freedom.
For the spherical fitness landscape the result is:
\begin{align}
f_{a, e}(Q)   
\propto \exp \left[-Q N/(4 Q_{tot})\right] \operatorname{K_0}\left[Q N/(4 Q_{tot})\right],\label{eqn:evo:fac}
\end{align}
where $K_0$ is the zero-order modified Bessel function of the second kind.
This is identical to Waxman and Welch's result for the model with maximum pleiotropy~\cite{bib:Waxman2005}.

\section{Extreme value theory for $\Delta$\label{sec:evo:appevt}}
The normalized spacing between cusps $\Delta$ is a ratio of two values; to calculate its probability density we first calculate the density of $i_1 \equiv \log \theta_1 - \log \theta_2$, the spacing between the logarithms of the largest two $\theta$s.
Defining
\begin{align}
\omega \equiv \log\left(\frac{\theta N}{Q_{tot}}\right)
\end{align}
and using the asymptotic $\chi^2$ density for $\theta$  yields the asymptotic probability density of $\omega$:
\begin{align}
f(\omega) = \exp{\left[-\frac{1}{2}\big(\exp\left(\omega\right) - \omega\big)\right]}/\sqrt{2 \pi}.
\end{align}
The corresponding cumulative probability distribution $F(\omega) \equiv \int_{-\infty}^\omega f(\omega') d\omega'$ is
\begin{equation}
F(\omega) = \erf\left(\exp\left(\omega/2\right)/\sqrt{2}\right),
\end{equation}
where $\erf$ is the error function.
This distribution has exponential-type extreme value statistics~\cite{bib:Gumbel1958}.

Following the terminology and notation of Gumbel~\cite{bib:Gumbel1958}, the typical size $u_{1,N}$ of the largest of N samples from the density $f(\omega)$ is given by $F(u_{1,N}) = 1-\frac{1}{N}$.
In our case this is
\begin{align}
u_{1,N} &= 2\log\left(\sqrt{2} \operatorname{erf}^{-1}\left(1 - 1/N\right)\right).\label{eqn:evo:u}
\end{align}
The corresponding scale parameter $\alpha_{1,N}$ is
\begin{align}
\alpha_{1,N} &= N f\left(u_{1,N}\right), \label{eqn:evo:alpha}
\end{align}
and distance $i_1$ between the largest two samples has probability density
\begin{align}
f(i_1) = \alpha_{1,N} \exp(-\alpha_{1,N}\, i_1).
\end{align}
(Gumbel's result~\cite{bib:Gumbel1958} for this distribution, his Eq.~5.3.5(4), has $\alpha_{2,N}$ in place of $\alpha_{1,N}$. In the limit $N\rightarrow\infty$ the two expressions are equal, but $\alpha_{1,N}$ is a better approximation for small $N$.)

The distance between the logarithms $i_1$ is related to $\Delta$ by $\Delta \equiv 1-\theta_2/\theta_1 = 1-\exp(-i_1)$.
Thus the probability density for $\Delta$ is
\begin{align}
f(\Delta) = \alpha_{1,N} \left(1-\Delta\right)^{(\alpha_{1,N} - 1)},
\end{align}
and the average of $\Delta$ is 
\begin{align}
\left<\Delta\right> = \frac{1}{1 + \alpha_{1,N}}.
\end{align}

A useful approximation for $\alpha_{1,N}$ can be obtained using an asymptotic expansion for $\erf^{-1}$~\cite{bib:Abramowitz1964}:
\begin{equation}
\sqrt{2} \operatorname{erf}^{-1}\left(1-x\right) \sim \sqrt{\log\left(\frac{2}{\pi x^2}\right) - \log\log\left(\frac{2}{\pi x^2}\right)}.
\end{equation}
Propagating this expansion through $u_{1,N}$ and $\alpha_{1,N}$ and neglecting terms of order $\log\log N$ in the final expression yields
\begin{equation}
\alpha_{1,N} \approx \log{N} + \log\left(\sqrt{2/\pi}\right).
\end{equation}
From this follows the approximate expression for $\left<\Delta\right>$ in Eq.~\ref{eqn:evo:approx}.

\section{Numerical simulation\label{sec:numerical_sampling}}

A random set of orthogonal unit vectors $\hat{v}_i$ can be obtained from the eigenvectors of a matrix $\mat{G}$ from the Gaussian Orthogonal Ensemble: $\mat{G} = \mat{H} + \transpose{\mat{H}}$ where the elements of $\mat{H}$ are standard normal random numbers.
A matrix $\mat{S}$ with eigenvalues $\lambda_i$ can then be constructed via
\begin{equation}
S_{j,k} = \sum_i \lambda_i v_{i,j} v_{i,k}.
\end{equation}

Random chemotypes $\vec{k}$ with specified log-fitness $Q_{tot} = -\log W(\vec{k})$ are obtained using the Cholesky decomposition $\mat{A}$ of $\mat{S}^{-1}$, defined by $\mat{A} \smalldot \transpose{\mat{A}} = \mat{S}^{-1}$. $\vec{k}$ is then given by
\begin{equation}
\vec{k} = \sqrt{2 Q_{tot}} \mat{A} \smalldot \hat{k},
\end{equation}
where $\hat{k}$ is a random unit vector.

\end{document}